\documentclass[a4paper,11pt]{article}
\pdfoutput=1 % if your are submitting a pdflatex (i.e. if you have
\usepackage{jcappub} % for details on the use of the package, please
\usepackage[T1]{fontenc} % if needed
\usepackage{journals}
\def \mpc {\mbox{\rm Mpc}}
\def \vel {\rm {km\,s^{-1}}}

\title{\boldmath CLASH-VLT: Testing the Nature of Gravity with Galaxy Cluster Mass Profiles}

\author[1]{L. Pizzuti,}
\author[1,2]{B. Sartoris,}
\author[1,2,3]{S. Borgani,}
\author[4]{ L. Amendola,}
\author[5]{K. Umetsu,}
\author[2]{A. Biviano,}
\author[1,2]{M. Girardi,}
\author[7]{P. Rosati,}
\author[2]{I. Balestra,}
\author[6]{G. B. Caminha,}
\author[11]{B.Frye,}
\author[10]{A. Koekemoer,}
\author[8]{C. Grillo,}
\author[12]{M. Lombardi,}
\author[9]{A. Mercurio,}
\author[2]{M. Nonino}

\affiliation[1]{Dipartimento di Fisica, Sezione di Astronomia, Universit\`a di Trieste,\\ Via Tiepolo 11, I-34143 Trieste, Italy}
\affiliation[2]{INAF - Osservatorio Astronomico di Trieste,\\ Via Tiepolo 11, I-34143 Trieste, Italy}
\affiliation[3]{ INFN - Sezione di Trieste,\\ Via Valerio 2, I-34127 Trieste, Italy}
\affiliation[4]{Institut f\"ur Theoretische Physik, Universit\"at Heidelberg,\\ Philosophenweg 16, D-69120 Heidelberg, Germany}
\affiliation[5]{5 Institute of Astronomy and Astrophysics, Academia Sinica,
P.O. Box 23-141,\\ Taipei 10617, Taiwan}
\affiliation[6]{Dipartimento di Fisica e Scienze della Terra, Universit\`a di
Ferrara,\\ Via Saragat 1, I-44122 Ferrara, Italy}
\affiliation[7]{University Observatory Munich, Scheinerstrasse 1, D-81679 Mu\"nchen, Germany}
\affiliation[8]{ Dark Cosmology Centre, Niels Bohr Institute, University of Copenhagen,\\ Juliane Maries Vej 30, DK-2100 Copenhagen, Denmark}
\affiliation[9]{Osservatorio Astronomico di Capodimonte, \\Via Moiariello
16, I-80131 Napoli, Italy}
\affiliation[10]{Space Telescope Science Institute, 3700 San Martin Drive,
Baltimore, MD 21208, USA}
\affiliation[11]{Steward Observatory/Department of Astronomy, University of Arizona, 933 N Cherry Ave, Tucson, AZ 85721,
USA}
\affiliation[12]{ Dipartimento di Fisica, Universit\`a degli Studi di Milano, via Celoria
16, I-20133 Milan, Italy}

\emailAdd{pizzuti@oats.inaf.it}
\emailAdd{barbara.sartoris@gmail.com}
\emailAdd{borgani@oats.inaf.it}
\emailAdd{l.amendola@thphys.uniheidelberg.de}

\abstract{We use high-precision kinematic and lensing measurements
of the total mass profile of the dynamically relaxed galaxy cluster 
MACS J1206.2-0847 at $z=0.44$ to estimate the value of the ratio
$\eta=\Psi/\Phi$ between the two scalar potentials in the linear
perturbed Friedmann-Lemaitre-Robertson-Walker metric. 
An accurate measurement of this ratio, called anisotropic stress, could show possible,
 interesting deviations from the predictions of the theory of General Relativity, according to which $\Psi$ should be equal to $\Phi$.
Complementary kinematic and lensing mass
profiles were derived from exhaustive analyses using the data from the Cluster
Lensing And Supernova survey with Hubble (CLASH) and the spectroscopic 
follow-up with the Very Large Telescope (CLASH-VLT).
Whereas the kinematic mass profile tracks only the time-time part
of the perturbed metric (i.e. only $\Phi$), the lensing mass profile
reflects the contribution of both time-time and space-space components
(i.e. the sum $\Phi+\Psi$). We thus express $\eta$ as a function
of the mass profiles and perform our analysis over the radial range
$0.5\,\mpc\le r\le r_{200}=1.96\,\mpc$. Using a
spherical Navarro-Frenk-White mass profile, which well fits the data, we obtain 
$\eta(r_{200})=1.01\,_{-0.28}^{+0.31}$ at the 68\% C.L. We discuss
the effect of assuming different functional forms for mass profiles
and of the orbit anisotropy in the kinematic reconstruction.
Interpreting this result within the well-studied $f(R)$ modified gravity model,
the constraint on $\eta$ translates into an upper bound to the interaction
length (inverse of the scalaron mass) smaller than 2 Mpc. This 
tight constraint on the $f(R)$ interaction range 
is however substantially relaxed when systematic uncertainties in
the analysis are considered. Our analysis highlights the potential
of this method to detect deviations from general relativity, while
calling for the need of further high-quality data on the total mass distribution of 
clusters and
improved control on systematic effects.}


\begin{document}
\maketitle
\flushbottom

\section{Introduction}

\label{sec:intro}
 In the current picture of cosmological studies,
the standard $\Lambda$CDM model seems to be the most suitable to
describe the expansion history, constrained with a wide range of observations
(e.g. ref. \cite{Planck13}). In this scenario, the nature of Dark Energy,
which is invoked to explain the current accelerating expansion (e.g. ref. \cite{Reiss01}) and
which should represent about 70\% of the total density of the Universe,
is still unknown. Several alternatives have been proposed
to explain cosmic acceleration, including the possibility of introducing
a modification of the Einstein's field theory of General Relativity
(hereafter GR), assumed in the Standard Cosmological model. These
models are expected not only to reproduce the expansion history, but
also to match general relativity at scales comparable with those of the Solar System,
where the theory is tested to high precision. Besides an accelerated
expansion, a modification of GR should manifest itself also through
its effect on the evolution of density perturbations. A possible evidence
of modified gravity should also involve a change into the relation
between the scalar potentials which appear in the perturbed Friedmann-Lemaitre-Robertson-Walker
(hereafter FLRW) metric and the fluctuations in the matter density field (ref. \cite{Lue01}).
Suitable parameterizations of these relations allow one to characterize
deviations from GR as a function of redshift and scale (e.g. see ref. \cite{PlanckMod}
and references therein).

In this paper we constrain deviations from GR by comparing
mass profiles of galaxy clusters as derived from gravitational lensing
and from a kinematic analysis of member galaxies. In fact, galaxies
moving within clusters under the action of gravity only feel the time-time
part of the perturbed FLRW metric, which is expressed by the potential
$\Phi$. On the other hand, the geodesics along which photons propagate
within clusters reflect the contribution of both time-time and space-space
components of the linear metric perturbations, i.e. they feel the
sum of the two potentials $\Phi+\Psi$. Since $\Phi=\Psi$ in standard
gravity with non-relativistic matter sources, mass profiles obtained from kinematic and lensing analyses,
under the assumption of GR, should coincide as long as GR itself is
valid. In other words, under the assumption that astrophysical and
observational systematics are well understood for both the kinematic
and lensing analyses, any deviation of mass profiles based on using
either photons or galaxies as tracers of the metric perturbations
should reflect a deviation from GR.

In the following, as a case study, we present results from MACS J1206.2-0847
(hereafter MACS 1206) a galaxy cluster at redshift $z=0.44$ 
for which
high-quality imaging and spectroscopic data have been analysed in
detail as part of the Cluster Lensing And Supernova survey with Hubble
(CLASH, ref. \cite{Postman01}) and the spectroscopic follow-up 
with the Very Large Telescope (CLASH-VLT, ref. \cite{Rosati1}) programs.
By using the reconstructed mass density profiles from kinematic analysis
of ref. \cite{Biviano01} and combined strong-weak lensing measurements of ref. \cite{UmetsuMACS},
we derive a relation between the mass profiles and the metric scalar
potentials $\Phi$ and $\Psi$ to estimate their ratio $\eta(r,z=0.44)$
under the assumption of spherical symmetry of the cluster mass distribution.
A similar analysis, that presented in ref. \cite{SartorisDM}, used
the mass profiles for the same cluster to obtain a constraint on the pressureless
nature of dark matter. The possibility of measuring $\eta$ from observations
and therefore to detect deviations from the GR was discussed in
ref. \cite{Amendola:2007rr,Jain01}, where it was pointed out that
by combining constraints on the metric potentials ratio $\eta$ and
on the evolution of density perturbations, it is in principle possible to distinguish
modifications of gravity from non-standard dark energy models.
In the following, we assume a flat
$\Lambda$CDM model with $\Omega_{m}=0.3$ for the matter density parameter
and $H_{0}=70$ km s$^{-1}$Mpc$^{-1}$ for the present-day Hubble
constant to convert observed angular scales into physical scales.

\section{Theoretical framework}

We describe the linear perturbation to the FLRW
metric associated to a galaxy cluster as (e.g. ref. \cite{MBW}) 
\begin{equation}
ds^{2}=a^{2}(\tau)\left\{ \left(1+2\frac{\phi}{c^{2}}\right)d\tau^{2}-2w_{i}d\tau dx^{i}-\left[\left(1-2\frac{\psi}{c^{2}}\right)\gamma(K)_{ij}+h_{ij}\right]dx^{i}dx^{j}\right\} ,
\end{equation}
where $w_{i}$, $h_{ij}$, $\phi$ and $\psi$ are functions of the
coordinates that characterize the perturbation, $\gamma(K)_{ij}$
is the three-dimensional metric tensor for a space with constant curvature
$K$. The general perturbed FLRW metric can be decomposed into scalar,
vector and tensor modes, by appropriate modifications of the functions
mentioned above. Since we are interested only in the scalar part of
the perturbation, related to the gravitational potential, we perform
a gauge transformation setting $w_{i}=h_{ij}=0$. The metric in spherical
coordinates takes now the form: 
\begin{equation} 
ds^{2}=a^{2}(\tau)\left\{ \left(1-2\frac{\Phi}{c^{2}}\right)d\tau^{2}-\left(1-2\frac{\Psi}{c^{2}}\right)[d\chi^{2}+f_{K}^{2}(\chi)d\Omega^{2}]\right\} .\label{eq:pertMetric}
\end{equation}
This gauge choice is called the conformal Newtonian gauge \cite{Mukhanov01},
for the similarity with the Newtonian limit of gravity. In the above
expression $\Phi$ and $\Psi$ are gauge-invariant scalar quantities,
the Bardeen potentials (see ref. \cite{Bardeen01}), with $\Phi$ playing
the role of the standard Newtonian potential. As usual, $a(\tau)$
is the expansion factor, which is a function of the conformal time $\tau$.
Finally, $f_{K}(\chi)$ is a function of the curvature $K$ and the
radial comoving coordinate $\chi$, with $f_{K}(\chi)=\chi$ for a
flat background universe with $K=0$.

The perturbed metric of eq. \ref{eq:pertMetric} is related to the
perturbation in the matter and energy content through the linearized
Einstein's field equations: 
\[
\delta G_{\mu\nu}=\frac{8\pi G}{c^{4}}\delta T_{\mu\nu}.
\]
If we assume the cosmological principle, the background energy momentum
tensor is that of an ideal fluid, 
\[
T^{\mu\nu}=\left(\frac{p}{c^{2}}+\rho\right)u^{\mu}u^{\nu}-g^{\mu\nu}\frac{p}{c^{2}},
\]
where $\rho$ and $p$ are the density and pressure of the fluid,
$u^{\mu}=dx^{\mu}/d\tau$ is the four-velocity. In this case, it can
be shown that the difference between the Bardeen potentials obeys
the following equation in Fourier space: 
\begin{equation}
k^{2}(\Psi-\Phi)=-8\pi Ga^{2}\bar{P}\Pi,
\end{equation}
with 
\[
\bar{P}\Pi=\frac{3}{2}\left(\frac{k_{i}k_{j}}{k^{2}}-\frac{1}{3}\delta_{ij}\right)\left({T^{i}}_{j}-\frac{1}{3}\delta_{ij}T\right)
\]
the stress tensor of the fluid. Since for an ideal fluid $\bar{P}\Pi=0$,
this leads to the condition that $\Phi=\Psi$. We thus define the anisotropic
stress $\eta(x)$ as the ratio $\Psi(x)/\Phi(x)$, where $x\equiv x^{\mu}$
are the spacetime coordinates; a deviation of this parameter from
the value $\eta=1$ indicates a violation of the Einstein's equations,
i.e.  a deviation from standard gravity.

In the following analysis, we carry out an observational determination
of $\eta$ by using the cluster mass profiles obtained from measurements
of the velocity dispersion of the cluster galaxies and from combined
strong and weak lensing measurements. These two methods to
infer mass profiles from observational data are connected to the gravitational
potentials in different ways. The motion of the galaxies in a cluster
is determined by the metric time-time component only $g_{00}=-(1+2\Phi/c^{2})$,
since their typical velocities, $\sim10^{3}{\,\rm km\, s^{-1}}$, are
non-relativistic. For example, in the case of the cluster MACS
1206, the velocity dispersion along the line of sight has been measured by ref. \cite{Biviano01}
$\sigma_{LOS}=1087_{-55}^{+53}\,\vel\ll c$. In the weak field limit
the geodesic equation, for a non-relativistic particle in a gravitational field reduces
to 
\[
\frac{d^{2}\vec{x}}{dt^{2}}=-\nabla\Phi\,.
\]
Here $\Phi$ is related to the source term by the (0,0) component
of the Einstein's equations, which in this context is simply given
by the Poisson equation 
\begin{equation}
\nabla^{2}\Phi=4\pi G\rho\,.\label{eq:dyn}
\end{equation}
On the other hand, photons perceive the gravitational field reflecting
the contribution of both time-time and space-space metric components.
This can be shown by using the general formalism of light propagation
in curved spacetime. Adopting an approach similar to Faber \& Visser
(ref. \cite{Faber06}), in a spherically symmetric, static spacetime
\footnote{In general the metric is also a function of the time coordinate, but
we can safely assume that the geometry does not change 
during the lensing and kinematic observations.
} (i.e. $\partial_{t}g_{00}=0$ and $g_{i0}=g_{0i}=0$) we define an
effective refractive index $n(r)$ that, in the weak field approximation,
is related to the perturbed metric coefficients as 
\begin{equation}
n(r)=\left[1-\frac{2}{c^{2}}\Phi_{lens}+O^{(2)}(\Phi,\Psi)\right],\label{eq:refractive}
\end{equation}
where we set $\Phi_{lens}=(\Phi+\Psi)/2$ to be the lensing potential.
Eq. \ref{eq:refractive} can be simply derived setting $ds^2=0$ in eq. \ref{eq:pertMetric}
and computing $c/v\equiv n$.

Using the lensing potential, we can define a lensing density field
$\rho_{lens}$ through the Poisson equation 
\begin{equation}
\nabla^{2}\Phi_{lens}=4\pi G\rho_{lens}\,.\label{eq:lens}
\end{equation}
Under the assumption of spherical symmetry one can integrate eq. \ref{eq:dyn}
and eq. \ref{eq:lens} over a sphere of radius $r$, thus obtaining
\begin{subequations} \label{eq:pot} 
\begin{equation}
\frac{d}{dr}\Phi(r)=\frac{G}{r^{2}}m_{dyn}(r),\label{eq:pot:1}
\end{equation}
\begin{equation}
\frac{d}{dr}\left[\Phi(r)+\Psi(r)\right]=\frac{2G}{r^{2}}m_{lens}(r)\,.\label{eq:pot:2}
\end{equation}
\end{subequations} In the above equations, $m_{dyn}(r)$ and $m_{lens}(r)$
are the total dynamic and lensing masses enclosed within a sphere
of radius $r$. Inserting eq. \ref{eq:pot:1} into eq. \ref{eq:pot:2},
we derive the relation between the ratio of
the Bardeen potentials and the cumulative mass profiles: 
\begin{equation}
\eta(r)\equiv\frac{\Psi(r)-\Psi(r_{0})}{\Phi(r)-\Phi(r_{0})}=\frac{\int_{r_{0}}^{r}\frac{G}{r'^{2}}\left[2m_{lens}(r')-m_{dyn}(r')\right]dr'}{\int_{r_{0}}^{r}\frac{G}{r'^{2}}m_{dyn}(r')dr'}.\label{eq:eta}
\end{equation}
Here $\Phi(r_{0})$ and $\Psi(r_{0})$ are two integration constants
that we can set equal to zero using the freedom in the definition
of the potentials.

\section{MACS 1206 and mass profiles}

\label{sec:3} The galaxy cluster MACS 1206, located at redshift $z=0.44$,
is one of the 25 X-ray selected clusters observed as part of
the CLASH survey. 

Specifically, twenty CLASH clusters, including MACS1206, have been selected 
from X-ray-based compilations 
of dynamically relaxed systems, and the remaining five 
clusters have been chosen for their lensing strength \citep{Postman01}. 
MACS1206 appears as a large-scale relaxed system 
with a few minor overdensities in 2D distribution \citep{girardi15}.
This result is also supported by the analysis of ref. \cite{Lemze01}, that
does not find a significant level of substructures within the cluster
when the most conservative selection is used to assign the membership
of cluster galaxies. 
The concentric distribution of the mass components (stellar,
gas and dark matter, see ref. \cite{UmetsuMACS}) further point to a relaxed status of the cluster 
(see also ref. \cite{hudson10} who demonstrate that
the projected separation of the
BCG and the X-ray emission peak is a robust indicator of a system's dynamical
state).
Moreover, the kinematic mass profile determination is in agreement
with the analysis based on the \emph{Chandra} X-ray observations under
the assumption of hydrostatic equilibrium. Mass profiles based on
hydrostatic equilibrium of the intra-cluster medium and on the Jeans'
equation are both sensitive to the time-time component of the metric,
but they feel the lack of equilibrium in different ways. Therefore,
the consistent results obtained by these methods suggest that the
cluster is in an equilibrium configuration (i.e. dynamically relaxed).

A first strong lensing analysis of MACS1206 was carried out
in ref. \cite{Zitirn01} using 50 multiple images of 13 background
lensed galaxies. An upgrade of this analysis was presented in ref.
\cite{UmetsuMACS}, exploiting a combination of strong lensing information
with weak lensing shear and magnification measurements from \emph{Subaru}
multi-band images. Taking advantage of high purity sample of background
galaxies derived from extensive multicolor and spectroscopic information,
a robust measurement of the cluster mass density profile was obtained
out to $\sim2\,\mpc$. In the radial region between $0.3$ and $0.4\,\mpc$,
the mass profiles derived from strong lensing and weak lensing shear
and magnification analyses have been shown to be consistent with each
other. The resulting mass profile is parametrized according to the
Navarro-Frenk-White model (NFW hereafter), ref. \cite{Navarro}, 
\begin{equation}
M(r)=M_{200}\frac{\ln(1+r/r_{-2})-(r/r_{-2})(1+r/r_{-2})^{-1}}{\ln(1+c)-c/(1+c)},\label{eq:nfw}
\end{equation}
where $M_{200}$ is the mass of a sphere with overdensity 200 times
the critical density of the universe at that redshift. Furthermore,
$r_{-2}$ is a scale radius that for the NFW model coincides with
the radius where the logarithmic derivative of
the mass density profile takes the value $d\ln\rho(r)/d\ln r=-2$.
Finally, the concentration parameter $c$ is defined as $c=r_{200}/r_{-2}$,
with $r_{200}$ the radius encompassing the mass $M_{200}$.

The measurement of the kinematic mass profile is presented in ref.
\cite{Biviano01} using spectroscopy information from the CLASH-VLT
project (ref. \cite{Rosati1}). Observations with VLT/VIMOS led to a total
of 2749 galaxies with reliable redshift measurements in the cluster field. After
the rejection of interlopers, 592 cluster members were identified.
The sample was then analyzed in the projected phase-space with the
\emph{MAMPOSSt} method of ref. \cite{Mamon01}, that solves the Jeans' equation
\begin{equation}
\frac{d(\nu\sigma_{r}^{2})}{dr}+2\beta(r)\frac{\nu\sigma_{r}^{2}}{r}=-\nu(r)\frac{G\, M(r)}{r^{2}}\,,\label{eq:Jeans}
\end{equation}
to provide a maximum likelihood fit for the parameters of different
mass models out to the virial radius ($\sim2\,\mpc$). In
eq. \ref{eq:Jeans} $\nu(r)$ is the galaxy number density and $\sigma_{r}$
indicates the radial velocity dispersion. The kinematic analysis also
requires modeling the velocity anisotropy profile $\beta(r)$ of the
tracers of the gravitational potential%
\footnote{The velocity anisotropy is defined as $\beta=1-(\sigma_{t}/\sigma_{r})^{2}$
where $\sigma_{t}$ and $\sigma_{r}$ are the tangential
and the radial
component of the velocity dispersion, respectively.%
}, due to the well-known mass-anisotropy degeneracy. In the original
analysis of ref. \cite{Biviano01} three models for $\beta(r)$ were
considered, specifically:  
\begin{equation}
\beta_{O}(r)=\beta_{\infty}\frac{r-r_{c}}{r+r_{c}},\label{eq:omod}
\end{equation}
\begin{equation}
\beta_{T}(r)=\beta_{\infty}\frac{r}{r+r_{c}},\label{eq:tmod}
\end{equation}
(from ref. \cite{Tiret01}), and constant anisotropy with no radial dependence
$\beta_{C}$. In the above equations, $\beta_{\infty}$ is the
anisotropy value at large radii. Finally, the parameter $r_{c}$ is
assumed to coincide with the radius $r_{-2}$ of the mass profile.
In fact, ref. \cite{Mamon10} proved that with this value of $r_{c}$ the
``O" and ``T" models provide a good fit to the average anisotropy profiles
predicted by a set of cosmological simulations of galaxy clusters.

The NFW model gives the highest likelihood fit to the kinematic data
for the mass profiles reconstructed with the MAMPOSSt method, and
in combination with the anisotropy \textquotedbl{}O\textquotedbl{}
model it gives the smallest product of the relative errors in the
two free parameters $r_{-2}$ and $r_{200}$. In our analysis, following
ref. \cite{Biviano01}, we adopt the combination of the NFW profile and the ``O''
model for the orbit anisotropy (``NFW+O" hereafter) as the reference
model, as obtained considering the scale radius of the number density profile $r_{\nu}=0.63$
 Mpc.
In order to estimate the dependence of the $\eta(r)$ measurement
(Eq. \ref{eq:eta}) on the kinematic mass profile used, we also
consider the Hernquist (ref. \cite{Hernquist01}, hereafter ``Her'' )
and the Burkert (ref. \cite{Burkert01}, hereafter  ``Bur") models, that provide
acceptable fits to the kinematic data (as shown in ref. \cite{Biviano01}).
In Table \ref{tab:i}, we summarize the kinematic mass models for which
we derived constraints on $\eta$.

\section{Results}

\label{sec:4} In this section we discuss the application of eq. \ref{eq:eta}
to compute the anisotropic stress $\eta(r)$ for MACS 1206, using
the lensing and kinematic mass profiles considered above. In our analysis,
we integrated the mass profiles in the radial range $[r_{0},r_{200}]$,
with $r_{0}=0.55\,\mpc$ and $r_{200}=1.96\, Mpc$, where the latter is the best-fit value
as obtained from both the kinematic and lensing analysis (refs. \cite{Biviano01,UmetsuMACS}).
At larger radii, dynamical equilibrium cannot be reliably assumed,
and therefore the Jeans' equation can no longer be used to infer the gravitational
potential. Moreover, at such large radii, the lensing masses become
less reliable, as the weak shear signal becomes increasingly contaminated
by large-scale structure filaments that might affect the recovered
mass profiles.

In spite of MACS 1206 global behavior (see Section \ref{sec:3}), 
in the innermost regions ($r\le0.5\,\mpc$) we do
not have sufficient information to establish whether the central core
is dynamically relaxed and to confirm the validity of the spherical
symmetry assumption. In fact, \textit{Chandra} X--ray observations
of MACS 1206 show an inner entropy profile (see ref. \cite{Cavagnolo01}) which
is higher than expected for a relaxed cool core cluster, thus indicating
a dynamically active core. Moreover, optical photometric observations,
presented in ref. \cite{Presotto01}, show evidence for a distribution of
the intracluster light (ICL) which is asymmetric with respect to the
position of the BCG, with an elongation in the direction of the second
brightest cluster galaxy. This suggests the presence of a tidal interaction
between these two central galaxies, further questioning dynamical
relaxation to hold in the central region of MACS 1206.

\begin{figure}
\includegraphics[clip,width=1\textwidth]{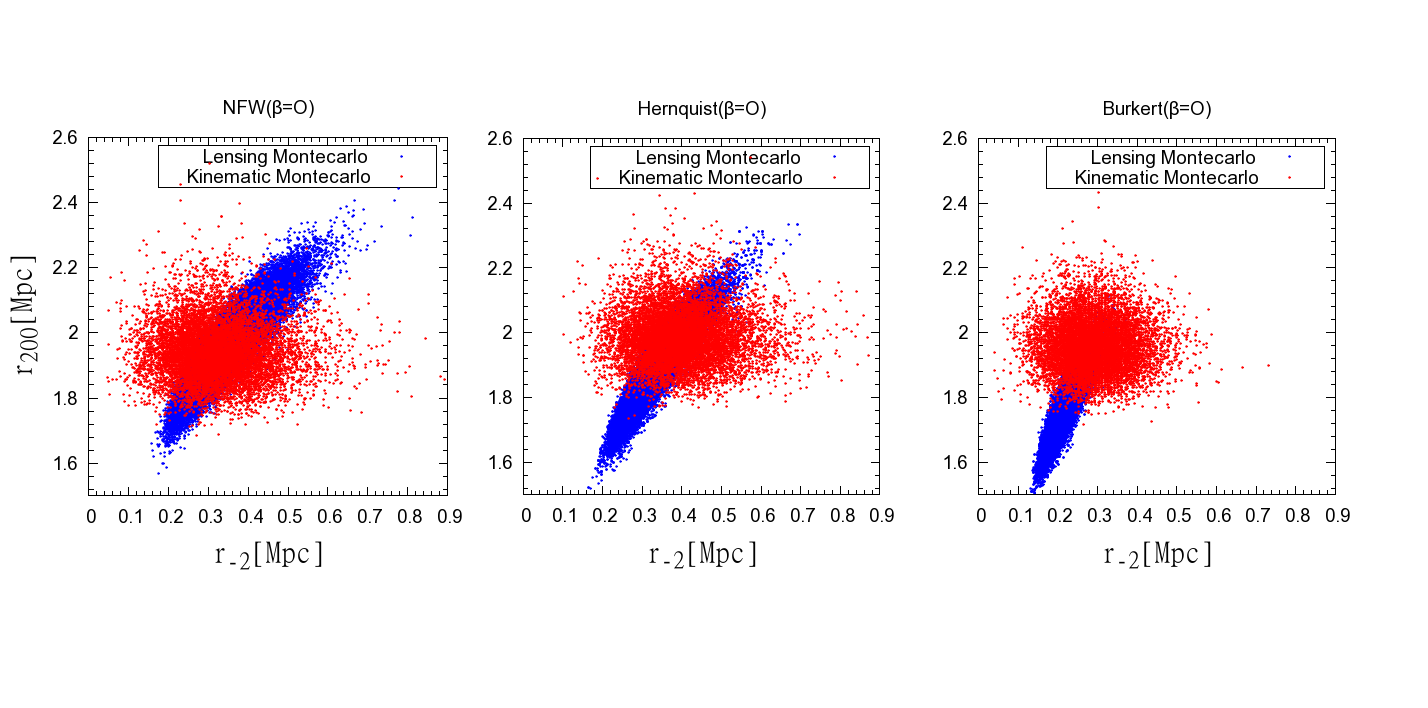}
 \caption{\label{fig:par} Results of the $10^{4}$ Montecarlo simulations generated
by sampling the joint probability distribution of $r_{200}$ and $r_{-2}$
from the kinematic analysis (red points) and from the combination
of strong and weak lensing analyses (blue points). Left panel: NFW
mass model of ref. \cite{Navarro}; central panel: Hernquist mass model
of ref. \cite{Hernquist01}; right panel:  Burkert mass model of ref. \cite{Burkert01}.}
\end{figure}

\begin{figure}
\centering
\includegraphics[clip,width=1\textwidth]{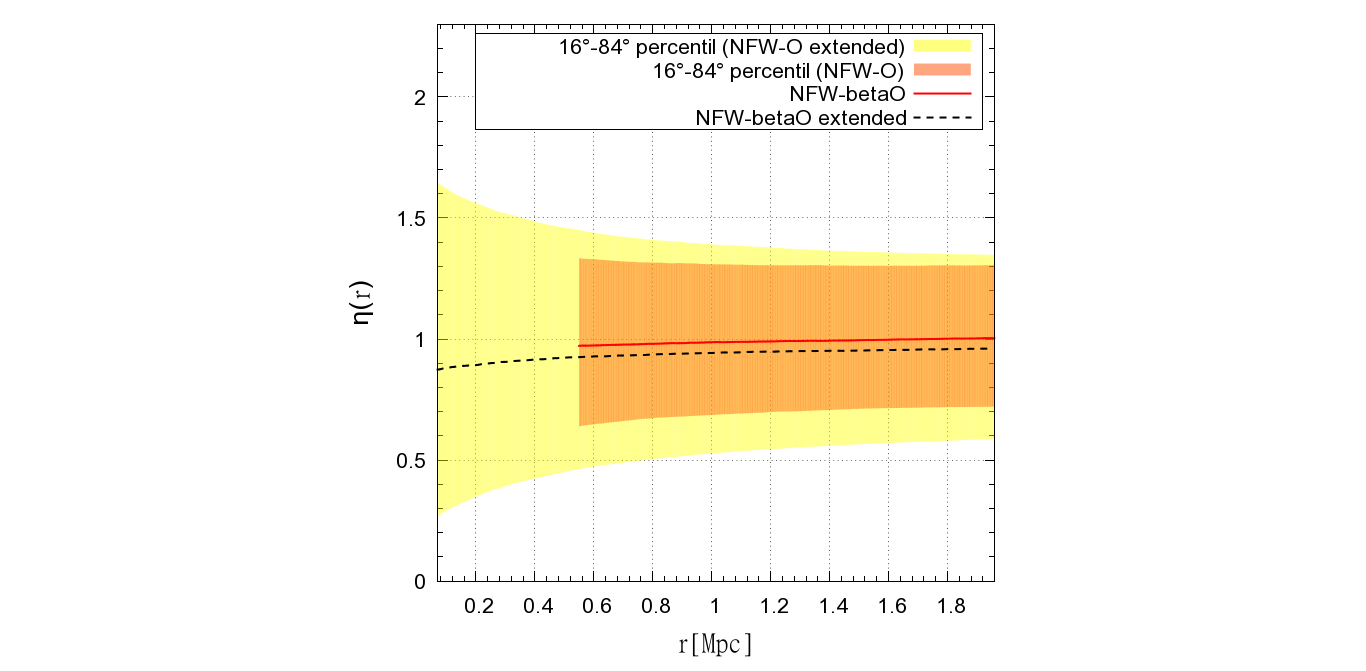}

\caption{\label{fig:eta3} Constraints on the radial profile of the anisotropic
stress $\eta(r)$ for the reference analysis based on the NFW parametrization
of the density profile (see eq. \ref{eq:nfw}) and the O-model of
eq. \ref{eq:omod} for the orbit anisotropy. Results correspond to
$r_{0}=0.55$ Mpc (NFW-betaO) and 0.07 Mpc (NFW-betaO extended)
 for the minimum radius down to which mass density profiles are considered. Solid red and black dashed
curves show the median values of $\eta(r)$, while the
narrower orange and broader yellow areas mark the corresponding 68 \% C.L. regions.}

\end{figure}

In our analysis we consider four parameters: $r_{-2}$ and $r_{200}$,
derived from the kinematic analysis in ref. \cite{Biviano01} and
$r_{-2}$ and $r_{200}$ derived from the lensing analysis by
ref. \cite{UmetsuMACS}.  In order to propagate the statistical errors
from the mass profile parameters to $\eta(r)$, and following the
analysis of ref. \cite{SartorisDM}, we repeated the calculation of
$\eta(r)$ by Montecarlo sampling with $10^{4}$ trials the two
probability distributions in the $(r_{-2},r_{200})$ parameter space,
as provided by the kinematic and by the lensing analyses.  The results
of these trials are shown in Figure \ref{fig:par} (from left to right:
NFW, Hernquist and Burkert mass profiles) with red and blue points for
the kinematic and lensing analysis, respectively.  As discussed in
ref. \cite{SartorisDM}, the joint distribution of the
$(r_{200},r_{-2})$ parameters from the kinematic analysis has nearly
zero covariance, so the errors on these two characteristic radii are
almost uncorrelated. On the other hand, the joint probability
distribution of the parameters from the lensing analysis can be
assumed to be a bi-variate Gaussian with covariance between $r_{-2}$
and $r_{200}$.  As expected, the iso-probability contours in the
$(r_{200},r_{-2})$ plane are almost elliptical in this case (see
Figure \ref{fig:par}).  In Figure \ref{fig:eta3} we plot our results
for $\eta(r)$ as a function of the distance from the center, $r$. For
our reference analysis based on the NFW+O mass model, we show the
results in the range $0.55-1.96\, $Mpc with the red solid line, along
with the corresponding 68\% C.L. (orange shaded area). The effect of
starting the integration of the mass profiles in eq. \ref{eq:eta} from
a smaller radius, namely $r_{0}=0.07$ Mpc, is shown by the black
dashed curve, with the yellow area marking the corresponding 68\%
C.L. 
The errors increase when we use information from the cluster central
region where the mass profile derived from strong-lensing has larger
errors than the one obtained from weak-lensing (see Figure 13 of ref. \cite{Biviano01}).
Here the errors in the strong-lensing regime are dominated by
model-dependent systematic uncertainties \cite{UmetsuMACS,zitrin15}.
The weaker
constraint affects all the $\eta$ profile by virtue of the correlation
between errors at different radii. We also notice that the median
values of $\eta(r)$ are slightly lower than those estimated when using
$r_{0}=0.55$ Mpc. In both cases, the results are consistent with
$\eta=1$, thus with the predictions of GR.

\begin{figure}
\includegraphics[width=\textwidth]{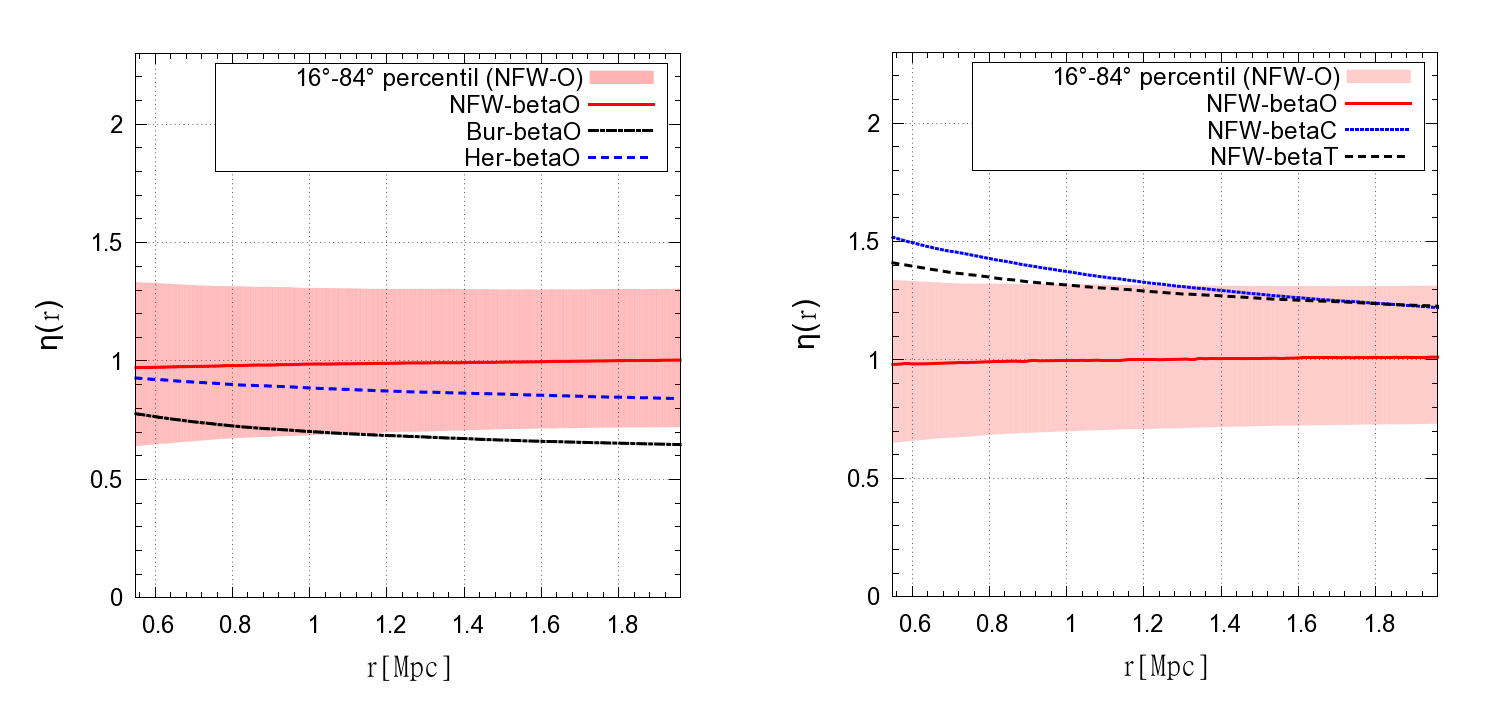}
\caption{\label{fig:eta2} The effect of changing the reference model for the
mass density profile (left panel) and the orbit anisotropy profile
$\beta(r)$ (right panel) on the resulting
constraints on the anisotropy stress profile $\eta(r)$. Here we assume
$r_{0}=0.55$ Mpc. Left panel: three mass profile models with fixed
anisotropy $\beta="O"$. NFW model, ref. \cite{Navarro}: red solid curve;
Hernquist model, ref. \cite{Hernquist01}: blue dashed curve; Burkert model,
ref. \cite{Burkert01}: black dash-dotted curve. 
Right panel: NFW mass profile for different
anisotropy models. $\beta="O"$: red solid curve; $\beta="T"$: black
dashed curve; $\beta="C"$: blue dotted curve. 
In both panels the shaded area indicates
the 68\% C.L. errors for the reference NFW+O model.}
\end{figure}

Possible systematic effects could in principle affect our
analysis. From an observational point of view, the cluster orientation
and asphericity can affect both lensing and kinematic mass profile
determinations. Ref.  \cite{girardi15} found that the ellipticity of
the galaxy distribution for MACS1206 is $\epsilon=
0.20^{+0.05}_{-0.06}$. Such value is low especially for a medium-$z$
cluster compared with what found at low redshift by ref. \cite{binggeli82}
($\left<\epsilon\right>= 0.25\pm0.12$ in a sample 44 Abell clusters)
and by ref. \cite{detheije95} ($\left<\epsilon\right> = 0.4$ for a sample
of 99 Abell clusters).  Moreover, since the kinematic analysis is
based on the Jeans equation and thus on the assumption that the
cluster is in dynamical equilibrium, the presence of substructures
could affect our results on $\eta$. However, as we discuss in Section
\ref{sec:3}, MACS1206 do not show a significant level of
substructures.

In Figure \ref{fig:eta2} we quantify the systematic effect on $\eta$
obtained by changing all the mass and anisotropy models.  We stress
that all the kinematic mass and anisotropy profile combinations
considered in this analysis have been proven in ref. \cite{Biviano01}
to provide acceptable fits and none of them is rejected by data. The
same finding also holds for the mass profiles from the lensing
analysis.  In the left panel, we consider three different mass
profiles (see also Table I) in both kinematic and lensing analysis,
assuming in all cases $\beta="O"$ for the orbit anisotropy.  The
solid, dashed and dash-dotted lines indicate the median values of the
distributions for NFW, Hernquist and Burkert respectively, while the
colored area indicates the  68\% C.L. region for the NFW+O profile. We note
that GR predictions are now slightly outside the 68\% C.L. regions
when using the ``Bur'' mass profile, thus underlining the importance
of the adopted mass profile parametrization.  However, we note that
noting that the ``Bur'' model has been statistically disfavored by the
ensemble mass profile derived from a stacked lensing analysis of the
CLASH X-ray-selected sample, also including MACS J1206, based on
strong--lensing, weak--lensing shear and magnification data
\cite{Umetsu15}.

In the right panel of Figure \ref{fig:eta2} we show
$\eta(r)$ computed using the NFW mass profile and the three anisotropy
profiles discussed in Section \ref{sec:3}. The shaded areas indicate the
68\% confidence regions for the reference model NFW+O while the solid, dashed and dotted
lines represent the medians of the distributions. In this case, although
the details of the results are sensitive to the anisotropy model adopted,
the resulting $\eta$ profiles always lie within the statistical uncertainties
of the reference model (see also Table I).

As mentioned above, since $\eta$ is obtained as a ratio of integrals
depending on the mass profiles (see eq. \ref{eq:eta}), the errors at different
radii are correlated. In fact, the determination of $\eta(r)$ at
a fixed radius $\bar{r}$ is affected by the shape of the profile
at $r<\bar{r}$. For this reason, we quote the values of $\eta$ computed
at $r_{200}=1.96\,$ Mpc for all the models analyses, as reported
in Table \ref{tab:i}.

\begin{table}
\centering %
\begin{tabular}{ccccc}
%\hline 
Mass profile  & $\beta$  & $\eta$  & $\Delta\eta(68\%C.L.)$  & $\Delta\eta(95\%C.L.)$\tabularnewline
%Mass profile  &  &  &  & \tabularnewline
 &  &  &  & \tabularnewline
\hline 
\hline 
 & $~~~~$  &  &  & \tabularnewline
NFW  & O  & 1.00  & $_{-0.28}^{+0.31}$  & $_{-0.54}^{+0.61}$\tabularnewline
 &  &  &  & \tabularnewline
NFW  & C  & 1.22  & $_{-0.38}^{+0.45}$  & $_{-0.68}^{+0.93}$\tabularnewline
 &  &  &  & \tabularnewline
NFW  & T  & 1.23  & $_{-0.33}^{+0.33}$  & $_{-0.61}^{+0.71}$\tabularnewline
 &  &  &  & \tabularnewline
Bur  & O  & 0.65  & $_{-0.23}^{+0.26}$  & $_{-0.44}^{+0.51}$\tabularnewline
 &  &  &  & \tabularnewline
Her  & O  & 0.84  & $_{-0.28}^{+0.31}$  & $_{-0.60}^{+0.66}$ \tabularnewline
 &  &  &  & \tabularnewline
\hline 
\end{tabular}\caption{\label{tab:i} Constraints on the anisotropic stress $\eta$ for the
different mass and anisotropy models. Column 1: mass model used
 (NFW: ref. \cite{Navarro}; Bur: ref. \cite{Burkert01};
Her: ref. \cite{Hernquist01}). Column 2: model for the profile of orbit
anisotropy as fitted in the kinematic mass reconstruction (O: eq.
\protect\ref{eq:omod}; T: eq. \protect\ref{eq:tmod}; C: constant
$\beta$). Column 3: median values for $\eta$ at $r_{200}$; Columns
4 and 5: errors at 68\% and 95\% C.L.}
\end{table}

For our reference model (NFW with anisotropy profile 'O'), we obtain
\begin{equation}
\eta(1.96\, Mpc)=1.00_{-0.28}^{+0.31}\,(\mbox{stat})\,\pm0.35\,(\mbox{syst}),\label{eq:eta_con}
\end{equation}
at 68 \% C.L. where the systematic error is computed taking into account the variation
in the median value of $\eta(r_{200})$, due to the different anisotropy
and mass profiles used. As such, our analysis provides constraints
on the anisotropic stress $\eta$ which are fully consistent with the
GR predictions. In order to highlight how these constraints compare
with those derived from other cosmological probes, in the following
section we will recast them in terms of constraints on the specific
class of $f(R)$ modified gravity models.

\section{Constraints on $f(R)$ models}

Constraints on $\eta$ can generally be used to set
bounds on specific modified gravity models. We consider here  for a moment
only the statistical error  given by eq. \ref{eq:eta_con}, and
assume therefore that $\eta$ lies between $0.7$ and $1.3$ at $r=1.96$
Mpc. Then we study how this translates into  constraints on one of the simplest
and most investigated class of modified gravity models, namely $f(R)$
models. In this class of models, proposed first in 1970 by ref. \cite{Buch01}
(see ref. \cite{DeFelice:2010aj} for a  review), the Einstein-Hilbert
Action 
\[
S_{EH}\,=\,\frac{1}{16\pi G}\int{\sqrt{-g}Rd^{4}x}
\]
is replaced by 
\begin{equation}
S\,=\,\frac{1}{16\pi G}\int{\sqrt{-g}[R+f(R)]d^{4}x},\label{eq:fr}
\end{equation}
where $f(R)$ is a function of the curvature scalar $R$. Varying
the action of eq. \ref{eq:fr} with respect to the metric $g_{\mu\nu}$
leads to the field equations: 
\begin{equation}
(1+f_{,R})R_{\mu\nu}-\frac{1}{2}g_{\mu\nu}[f(R)+R]+(g_{\mu\nu}\Box-\nabla_{\mu}\nabla_{\nu})f_{,R}=8\pi GT_{\mu\nu},\label{field}
\end{equation}
where $f_{,R}=df(R)/dR$.

The general expression for the two Newtonian potentials in $f(R)$
models in Fourier space is 
\begin{align}
 & \Phi=-\frac{4\pi G\rho_{m}}{1+f_{,R}}\frac{a^{2}}{k^{2}}\left(1+\frac{1}{3}\frac{k^{2}}{M^{2}a^{2}+k^{2}}\right),\\
 & \Psi=-\frac{4\pi G\rho_{m}}{1+f_{,R}}\frac{a^{2}}{k^{2}}\left(1-\frac{1}{3}\frac{k^{2}}{M^{2}a^{2}+k^{2}}\right).
\end{align}
where $k$ is the norm of the comoving wavevector. In the above expressions
for the potentials, the effective scalaron mass 
\begin{equation}
M^{2}=\frac{R}{3}\left(\frac{1+f_{,R}}{Rf_{,RR}}-1\right).
\end{equation}
(where $f_{,RR}=d^{2}f(R)/dR^{2}$) provides the interaction range
$\lambda=M^{-1}$ in the corresponding Yukawa-type potential. These
expressions are valid provided there are no screening mechanisms active
at the relevant scales. In standard GR $f_{,R}=0,\,\, f_{,RR}=0$
and $M\to\infty$, so that one recovers the usual expression for $\Psi,\Phi$
and for $\eta$. The expression for $\eta(r)=\Psi(r)/\Phi(r)$ is
the ratio of the Fourier anti-transforms of the potentials, and one
could derive directly constraints on $M$ and $f_{,R}$ at the scale
and redshift of the cluster. Since the present constraints are still
dominated by systematics is probably not worth to try a very detailed
comparison with theory. Therefore, our aim here is just to point out
the potential of this method. In this spirit, we derive here below
constraints on the parameters that characterize a $f(R)$ model.

In the limit of small scales, $k\gg M$, one has simply 
\begin{equation}
\eta=\frac{\Psi}{\Phi}=\frac{1}{2}
\end{equation}
while in the opposite limit $k\ll M$ one recovers the standard GR
result $\eta=1$. It is then clear that if we can rule out $\eta=1/2$,
which indeed appears at almost 2$\sigma$ from our best fit when considering
only the statistical error, we can say that the interaction scale
$\lambda=M^{-1}$ should be smaller than the cluster scale, i.e. 
\begin{equation}
\lambda<2\,\,\mathrm{Mpc}.
\end{equation}
In $f(R)$ models the first derivative $f_{,R}$ is already well constrained
to be much smaller than unity in order to pass cosmological constraints
(see, e.g., ref. \cite{PlanckMod}, and references therein); moreover,
for small deviations from standard gravity $f_{,RR}\ll R^{-1}$ so
we can approximate $M^{2}=1/(3f_{,RR})$ and we obtain at $z=0.44$
\begin{equation}
|f_{,RR(0.44)}|<1.3\,\,\mpc^{2}.
\end{equation}
This is the first time a constraint on the second derivative
of $f(R)$ is obtained without relying on a particular $f(R)$ model.
However, we emphasize again that we are neglecting the systematic errors,
so this constraint should be taken with some caution.

If the $f(R)$ model can be approximated near the present epoch by
a power law $f(R)=\alpha R_{0}(R/R_{0})^{-n}$ where $R_{0}\approx H_{0}^{2}$
is the present curvature scalar, then we find for the dimensionless
constants $\alpha,n$ the constraint 
\begin{equation}
|n(n+1)\alpha|=|(n+1)f_{,R0}|<\left(1.3H_{0}^{2}\right)\approx10^{-7}.
\end{equation}
(we are neglecting here the evolution between $z=0.44$ and the present
time). This is very close to, or better than, the current constraint $|f_{,R0}|=|\alpha n|\le10^{-6}$
(see e.g. refs. \cite{Xu01,Cataneo}) for compatibility with background
and linear perturbation theory. Of course the $f(R)$ Lagrangian can
have any shape so there is no need to expect necessarily a simple
connection between $f_{,R}$ and $f_{,RR}$. In general, the power
of estimating $\eta$ from lensing and kinematic mass profiles of
clusters is that one can put an independent constraint also on the
second, rather than just the first, derivative of $f(R)$ at scales
of order of one Mpc, much smaller than what is obtainable from
linear perturbation theory. 
Moreover, such constraints can be obtained independently at various
redshifts and do not require a specific $f(R)$ model valid along
the entire cosmic evolution which is instead needed when constraining
$f(R)$ theories with the cosmic microwave background or the linear
perturbation growth.

\section{Conclusions}

In this paper we have presented a method to derive constraints on the anisotropic
stress $\eta=\Phi/\Psi$ by comparing high-precision determinations
of the total mass profiles of galaxy clusters from lensing and kinematic analyses.
As a case study, we have applied this method to MACS 1206, a \textit{bona
fide} relaxed cluster at $z=0.44$ with $M_{200}=1.4\pm0.2 M_\odot$ (ref.\cite{Biviano01}). 
Lensing masses for MACS 1206 have been derived
by ref. \cite{UmetsuMACS} using the high-quality imaging and photometric
data obtained from HST and \textit{Subaru} within the CLASH project.
Kinematic mass profiles have been derived by ref. \cite{Biviano01} thanks
to intensive spectroscopic observations carried out within the CLASH-VLT
program. Galaxy motions are sensitive only to the time-time component
$\Phi$ of the metric perturbation, while lensing is sensitive to
both the time-time and space-space components, i.e. to $\Phi+\Psi$.
Therefore, a comparison of mass profiles based on these two independent
methods allows one to set constraints on possible deviations from
the prediction of General Relativity (GR), $\eta=1$.

The results of our analysis can be summarized as follows. 
\begin{itemize}
\item Comparing mass profiles over the range of radii from $r_{0}=0.55$
Mpc out to $r_{200}=1.96$ Mpc, we find results to be consistent with
the prediction of GR: 
$\eta(r_{200})=1.00_{-0.28}^{+0.31}$ at 68\% C.L. for
our reference analysis based on the NFW parametrization of
the mass density profile and a specific model for the profile of orbit
anisotropy. 
\item While the above errors refer only to statistical uncertainties, we
also estimated the effects of systematic uncertainties related to
changing the parametrization of the mass density and orbit anisotropy
profile, as well as changing the minimum radius down to which mass
profiles are considered. Within the range of models considered, 
these systematic uncertainties roughly double
the uncertainty in the measurement of $\eta$. 
\item To illustrate the potential of the method, we re-phrased the
  constraints on $\eta$, with a 30\% uncertainty, in terms of
  constraints on the $f(R)$ class of modified gravity models,
  neglecting systematics.  Interestingly, we find these constraints to
  be competitive with those obtained by combining expansion probes,
  cosmic microwave background anisotropies and large-scale structure
  observations (e.g., refs. \cite{Xu01,PlanckMod,Cataneo}).  In
  particular, ref. \cite{Zhao10} constrained the values of $\eta(z=0)$
  and $\mu(z=0)$, where $\mu(z,k)$ represents the modification to the
  Poisson equation for $\Phi$, according to
\begin{equation}
 -k^2\Phi=4\pi Ga^2\mu(z,k)\delta\rho,
\end{equation}
with $\delta\rho$ the overdensity of the perturbation. This analysis
was based on WMAP-5 data combined with cosmic shear data from CFHTLenS
and Integrated Sachs Wolf (ISW) data, taking into account also a
possible time evolution of the two functions. They found
$\eta(z=0)=0.98^{+0.73}_{-1.00}$ for $z_s=1.0$ and $\eta(z=0)=1.30\pm
0.35$ for $z_s=2$, where $z_s$ is a transition redshift at which the
parameters smoothly change to their late time values, with
uncertainties referring to 68\% C.L..  In a similar way,
ref. \cite{Daniel10} obtained $-1.6<\eta(0)-1<2.7$ at 95\% C.L. by
combining CMB constraints from WMAP-5, Type-Ia SN from Union2, and
cosmic shear data from CFHTLS and COSMOS surveys.

Ref. \cite{PlanckMod} combined CMB data with different cosmological
probes to study time- and scale-dependence of the modified gravity
parameters $\eta$ and $\mu$, both extrapolated to $z=0$.  As a result,
a tension with GR is found at $\sim 3\sigma$ C.L.  when \emph{Planck}
CMB data are combined with constraints from Baryonic Acoustic
Oscillations, Redshift Space Distortions and Weak Lensing
data. Interestingly, the tension with $\Lambda$CDM predictions is
alleviated when including in the analysis also the contribution from
CMB lensing (see Table 6 and also Figures 14,17 of
ref. \cite{PlanckMod}).  In this case they obtain $\eta(z=0)-1=0.60\pm
0.86$ for the scale-independent determinations. They also show that
the constraints become weaker when introducing the dependence on the
scale. 

In general, our
results are broadly consistent with the above constraints on
deviations from GR, even if our method provides constraints only on $\eta$.
Our statistical uncertainty in the measurement of $\eta$ is quite
competitive with those obtained from CMB and large-scale structure
probes. However, we emphasize once again that an accurate control of
the systematics in our analysis is mandatory for our
proof-of-concept analysis to turn into an accurate and robust method
to constrain modifications of gravity at the scales of galaxy
clusters. 
\end{itemize}
It is worth pointing out that the above results have been obtained
from high-quality observational data of only a single galaxy cluster,
thus highlighting the potential of using mass profiles of clusters
as tools to probe the nature of gravity on cosmological scales. In
principle, this result should not surprise; as long as the cluster
in consideration satisfies the main assumptions on which lensing and
kinematic mass profiles are recovered, the precision of the derived
constraints is only limited by the quality of observational data.
Kinematic mass profiles are based on solving the Jeans equation for
the projected phase-space distribution of cluster galaxies, assuming
a spherically symmetric stationary system within which galaxies moves
as tracers of the underlying potential. Even though the lensing mass
profile do not rely on any assumption on the dynamical state of the
cluster, its reconstruction still assumes spherical symmetry, as well
as negligible contamination from the surrounding large-scale structure.
In this respect, the choice of MACSJ 1206 for this case study is close
to be optimal, given the overall appearance of this object as dynamically relaxed system.

MACSJ 1206 is only one of a dozen clusters of the CLASH-VLT survey for
which data of comparable quality are available.  
The extension of this analysis to other clusters requires the
combination of large redshift samples, high-quality weak and strong lensing
data, as well as X-ray data on well selected clusters. A sample of at
least 500 redshifts of member galaxies
is needed for accurate dynamical mass profiles. Together with
deep X-ray data, kinematic data are also needed to check whether the system is
relaxed or whether other astrophysical systematics can play a significant
role. As a complementary approach, applying the same
analyses to realistic cosmological simulations of galaxy clusters
should quantify the impact of systematics in the measurement of
lensing and kinematic mass profiles, and, ultimately, their impact
in precision tests of gravity at the scale of galaxy clusters.\\

\noindent \textbf{Acknowledgements.} 
L.P. and S.B. acknowledge support from the PRIN-MIUR 201278X4FL grant
and from the ``InDark'' INFN Grant.  B.S., A.B. and M.G. acknowledge
support from the PRIN MIUR 2010-2011 (J91J12000450001) grant and and
from PRIN-INAF 2014 1.05.01.94.02.  L.A. is supported by the DFG TR33
``The Dark Universe'' grant.  K.U. is supported by the grants MOST
103-2112-M-001-003-MY3 and MOST 103-2112-M-001-030-MY3.
C.G. acknowledges support by VILLUM FONDEN Young Investigator
Programme through grant no. 10123. G.B.C. is supported by the CAPES-ICRANET programme through the grant
BEX 13946/13-7. This work is partially supported by
``Consorzio per la Fisica di Trieste''.

 \bibliographystyle{JHEP}
\bibliography{master}

\end{document}